\documentclass[english,aps,pra]{revtex4}
\usepackage[T1]{fontenc}
\usepackage[latin1]{inputenc}
\usepackage{subfigure}
\usepackage{graphicx}
\usepackage{amssymb}
\usepackage{amsmath}

\makeatletter

\usepackage{babel}
\makeatother
\begin{document}

\title{Dynamics of Atom-Field Entanglement from Exact Solutions: \\ Towards Strong-Coupling
  and Non-Markovian Regimes}

\author{Nick Cummings}

\email{nickc@umd.edu}

\affiliation{Joint Quantum Institute and Department of Physics,\\
University of Maryland, College Park, Maryland 20740}

\author{B. L. Hu}

\email{blhu@umd.edu}

\affiliation{Joint Quantum Institute and Department of Physics,\\
University of Maryland, College Park, Maryland 20740}

\date{August 16, 2007}

\begin{abstract}
  We examine the dynamics of bipartite entanglement between a two-level atom and the
  electromagnetic field. We treat the Jaynes-Cummings model with a single field mode
  and  examine in detail the exact time evolution of entanglement,
  including cases where the atomic state is initially mixed and the atomic transition
  is detuned from resonance.  We then explore the effects of other nearby
  modes by calculating the exact time evolution of entanglement in more complex systems
  with two, three, and five field modes. For these cases we can obtain exact solutions which
  include the strong-coupling regimes. Finally, we consider the entanglement of a two-level atom with the infinite collection of modes present in the intracavity field of a Fabre-Perot cavity.
  In contrast to the usual treatment of atom-field interactions with a continuum of modes
  using the Born-Markov approximation, our treatment in all cases describes the full non-Markovian dynamics of
  the atomic subsystem. Only when an analytic expression for the infinite mode case
  is desired do we need to make a weak coupling assumption which at long times approximates 
  Markovian dynamics.
\end{abstract}
\maketitle

\section{Introduction}

In recent years the quest for a quantum computer has generated much
interest in quantum information processing \cite{NielsenChuang}
and the need for a deeper understanding of the distinguishing quantum
features of many body systems, such as quantum coherence and
entanglement. Recent progress in many areas of physics, especially
atomic-optical, condensed matter, low temperature and mesoscopic
physics, has been brought about by new techniques (in manipulating
atoms and light) and high precision measurements and fabrication of
new materials. The drive for better preparation, manipulation
and control on the experimental side is matched by an adjustment of
focus and emphasis of theoretical studies. There is a trend to go back
to simpler models (e.g., Bose-Hubbard for BEC) but explore the
parameter regimes and physics hitherto unknown or unclear (such as the
BCS-BEC transition, BEC-Mott transition) but made possible only
recently by the high precision and highly manipulative experiments.
With these efforts new demands on experimental designs (such as
accurately positioning an atom in a cavity \cite{Nussmann,Lynn}) for
new devices (such as single atom laser \cite{KimbleSAL,SunYou}) and
new theoretical issues (such as quantum fluctuations and correlations,
decoherence and disentanglement) come to the fore.

It is in this spirit that we explore some of these newly introduced
issues in atom-field interaction, with the stated purpose of better
understanding the effects of non-Markovian dynamics and strong
coupling between a two-level atom (2LA) and the electromagnetic field
(EMF).  In this paper we will seek first to give a detailed, general, 
and exact treatment of the dynamics of entanglement in the 
Jaynes-Cummings model.  From there, we expand our considerations to 
include the effects of other, nearby field modes, which can, in principle, 
add new features to the evolution of entanglement when coupling is strong 
enough.  Finally, we will explore interaction between the atom and the full, 
infinite collection of modes in the intracavity field. In all these 
cases we solve for the exact dynamics
\protect \footnote{Exact up to the approximations 
inherent in the usual model for the Hamiltonian of the system, which we will 
discuss} 
keeping the full unitary dynamics of the composite system, so that 
our treatment can capture the non-Markovian behavior of the atom
(considered as a subsystem) even in the infinite-mode case.  In this
approach, then, we 
wish to give a detailed description of the much-studied, single-field-mode Jaynes-Cummings 
model and connect this to strong-coupling and non-Markovian phenomena in more general models. 
We want to adopt an approach which can best preserve the quantum coherence and entanglement of the
system and include the full interplay of the subsystems involved (or
back-action from its environment, if any one such subsystem merits
special attention), treated self-consistently. The
influence functional approach we take here is particularly suitable
for such requirements, and they are necessary for a correct treatment of
disentanglement, because in general it is an even more intricate and
delicate issue than decoherence. 

Simpler atom-field models, such as that of a 2LA interacting with one
single EMF mode described by the Jaynes-Cummings model, have
solutions in closed form. Such closed form solutions offer a good comparison with results
involving approximations.  Quantum entanglement shows up as `collapse
and revival' phenomena, and has been studied in depth by Pheonix and
Knight \cite{PhoKni,Buzek} and Gea-Banacloche \cite{Gea90} 15 years
ago for initial pure states and Bose et al \cite{Bose} for initial
mixed states. We analyze this problem to obtain the exact time
evolution of the entanglement (rather than just a bound on or average of 
entanglement) for general, mixed initial atomic states, and we explore the effect of detuning
between the atomic transition frequency and the cavity mode
frequency.   This gives us a detailed and exact picture of the dynamics 
of entanglement in the well-known Jaynes-Cummings model. 

 Moving on to more complex models, we study the case of a 2LA interacting 
with two, three, or five EMF modes.  We present an exact solution of the 
time evolution of entanglement showing the effect of these additional models for an arbitrary, pure 
initial atomic state.  Using a perturbative approach we give leading order 
corrections to Jaynes-Cummings dynamics from other, nearby field modes in the 
case of strong atom-field coupling. There is a recent proposal to use the 2LA as a control qubit
and the two EMF modes as a target qubit in the making of a CNOT gate
\cite{CNOT}. Our calculation of entanglement between an atom 
with an initial mixed state and multiple field modes may be 
useful for these proposed studies. 

In support of the new needs mentioned above, our work aims at
probing the strong-coupling regime. Strong coupling is defined as $g
>> \kappa,\gamma$, where $\kappa$ is the cavity damping rate and
$\gamma$ is the rate of spontaneous emission from the atom. In this regime the coherent evolution of the atomic and
intracavity field state is fast compared to the dissipation rates in
the system, so coherent quantum effects become important
\cite{StrongCplg}. This is the regime we explore in treating the few
mode cases.

Beyond the strong-coupling regime,
there is also the regime of super-strong coupling
$g \simeq\Delta$, where $\Delta$ denotes free spectral range of the
resonator in angular frequency. While, conventionally, strong coupling is achieved through
recycling of the light by means of a high Q cavity, in the super
strong regime the coupling between atoms and light is already strong
during one round trip in the resonator.  Thus the atom can affect the
mode structure, in addition to the occupation of the modes, as is the
case in conventional Cavity QED.  (This, in our language is when the
back-action of the atom on the field begins to become important.)
According to \cite{MeiMey}, this super-strong-coupling regime can be
achieved for a modest number of ultracold atoms in an optical lattice
formed by standing waves in an optical resonator.  Our studies of cases with a few field modes cover
this super-strong regime for the case of a single atom, showing new
behavior in this regime, and they apply to certain situations for the multi-atom scenario which is experimentally realistic.

Finally, we study the
interaction of a 2LA with a cavity and its full, infinite collection
of EMF modes in certain limits, going beyond a simple one- or few-mode
treatment. We do not need to invoke 
the Born-Markov approximation and our approach generally applies to 
strong-coupling regimes.  Our calculations yield formal solutions for the 
evolution of entanglement in terms of integrals, which admit numerical solutions in the 
general case, but in order to obtain simple analytic solutions we explore a solution 
for late-time behavior that is perturbative in the atom-field coupling, and, so, those 
expressions are limited to the weak-coupling regime of cavity QED and exhibit 
behavior corresponding to Markovian expectations. 

The second aim for our work is to probe into the non-Markovian
regimes. Since our system, comprised of one atom together with the modes of the EMF, is closed, the dynamics of the complete system is
unitary. One may wonder where non-Markovian dynamics arises. This is a
reasonable question because non-Markovian behavior is usually
associated with open system dynamics, which is the case when one
focuses on one of the subsystems, often the 2LA, and introduces some
coarse-graining in the other subsystem, such as EMF, whence one begins
to see dissipative and stochastic behavior in the open system.
However, strictly speaking, non-Markovian just refers to
history-dependent processes (involving memories), and it is not just
for open systems, in the following sense: It is common knowledge in
non-equilibrium statistical mechanics \cite{Zwanzig} that for two
interacting subsystems the two ordinary differential equations
governing each subsystem can be written as an integro-differential
equation governing one such subsystem, thus rendering its dynamics
non-Markovian, with the memory of the other subsystem's dynamics
registered in the non-local kernels. All these are happening in a
closed system except now one has shifted the attention to one of its
subsystems.
\protect \footnote{Should the other subsystem possess a
  much greater number of degrees of freedom (called environment) and
  are coarse-grained in some way, the non local kernels are then
  responsible for the appearance of dissipation and noise. It is the
  act of coarse-graining which turns this particular subsystem of
  interest open.}
When an exact solution for the combined system is
available, as in many of the cases studied here, one can derive the
non-Markovian dynamics in the corresponding cases when the above
mentioned procedures are taken, or if the non-Markovian dynamics is
available from an open system consideration, one can identify the
coarse graining measures introduced. The commonly invoked Markovian
approximation gives rise to dynamics farther remote from (meaning,
less accurate than) the exact solutions.

We must add that investigations into the strong-coupling and
non-Markovian regimes are two rather difficult tasks, even for cavity
QED conditions. This study takes only a small step toward them. Our
aim is to gain some intuitive understanding of the evolution of
quantum entanglement in these new territories with a familiar and
relatively solvable model.  Much more work is needed to understand
the general features for more complicated systems in these regimes.

In Sect. II we specify the details of the physical system we wish to
treat and discuss how we will quantify the entanglement present.
Sect. III discusses the method we adapted to find the evolution of
the system and the effects of initial conditions assumed. We then
treat the specific cases of different collections of EMF modes in
Sect. IV, including a field consisting of a single mode, a few modes,
and the full collection of modes present in an idealized optical
cavity. We end in Sec. V with a discussion and a summary.

\section{System and Quantities Under Investigation}

\subsection{Hamiltonian and Initial States Considered}

The system we consider is a two-level atom (2LA) coupled to M
electromagnetic field (EMF) modes. We adopt the dipole and rotating-wave
approximations, neglecting motion of the atomic center of mass. The
Hamiltonian of this system, as given by \cite{AH}, is
\begin{equation}
  \hat{H}=\hbar\omega_{0}\hat{S}^{+}\hat{S}^{-}+\sum_{k=1}^{M}\hbar\omega_{k}\hat{b}^{\dagger}\hat{b}+\hbar(g_{k}\hat{S}^{+}\hat{b}+g_{k}^{*}\hat{S}^{-}\hat{b}^{\dagger})\equiv\hat{H}_{0}+\hat{H}_{I}\label{eq:totalham}\end{equation}
where
\begin{equation}
  \hat{H}_{0}=\hbar\omega_{0}\hat{S}^{+}\hat{S}^{-}+\sum_{k=1}^{M}\hbar\omega_{0}\hat{b}^{\dagger}\hat{b}\label{eq:freeham}\end{equation}
is the free Hamiltonian and
\begin{equation}
  \hat{H}_{I}=\sum_{k=1}^{M}\hbar\delta_{k}\hat{b}^{\dagger}\hat{b}+\hbar(g_{k}\hat{S}^{+}\hat{b}+g_{k}^{*}\hat{S}^{-}\hat{b}^{\dagger})\label{eq:interactionham}\end{equation}
is the interaction Hamiltonian. Here $\omega_{0}$ is the (bare)
frequency of the atomic transition, $\omega_{k}$ is the frequency of
the $k^{th}$ field mode, $g_{k}$ is the coupling of the $k^{th}$ mode
to the atom, and $\delta_{k}\equiv\omega_{k}-\omega_{0}$. Since
$g_{k}\propto1/\sqrt{\omega_{k}}$ implies
$g_{j}=g_{k}\sqrt{\frac{\omega_{k}}{\omega_{j}}}$, we may choose to
define things such that all values $g_{k}$ are real. Where the atom
and the electromagnetic field modes in question form a closed system,
the dynamics of the total system is unitary. For the $M=1$ case, this
is just the familiar Jaynes-Cummings model with no dissipation.

We will assume a separable initial state described by the density
matrix\begin{equation}
  \hat{\chi}(0)=\hat{\rho}_{A}(0)\otimes|0\rangle\langle0|\label{eq:init}\end{equation}
where all electromagnetic field modes are in the vacuum state
$\left|0\right\rangle $ and the atom may be in an arbitrary
(possibly mixed) state, described by the atomic reduced density matrix
$\hat{\rho}_{A}(0)$. If $\hat{\rho}_{A}(0)$ is pure then
$\hat{\chi}(t)$ is pure for all times due to the unitary evolution.

Tracing over the field, the reduced density matrix of the 2LA at
time $t$ is given by $\hat{\rho}_{A}(t)\equiv Tr_{F}(\hat{\chi}(t))$.
As is well known, it can be represented by a point in the Bloch
sphere. We will use the spherical polar parametrization of the sphere,
so that $\hat{\rho}_{A}$ can be specified by the triple
$(r,\theta,\phi)$, and, as usual, pure states lie on the surface of
the sphere with $r=1$. We choose $(1,0,\phi)$ to represent the excited
state of the atom $|e\rangle$ and $(1,\pi,\phi)$ to represent the
ground state $|g\rangle$.

In this work we seek to quantify the bipartite entanglement between
the atom and the electromagnetic field (considered as a whole). We
will not look at entanglement with individual modes of the field
separately, nor will we explore entanglement among the modes of the
field. A local unitary operation of the form
$\hat{U}=\hat{U}_{Atom}\otimes\hat{U}_{Field}$ does not change the
amount of entanglement in the system \cite{EntIntroPV}, and the time
evolution due to $\hat{H}_{0}$, which relates the Schr\"{o}dinger
picture to the interaction picture, is such a local unitary
operation; therefore, we may work entirely in the interaction picture
and compute the entanglement of the interaction picture state
directly, as though it were the Schr\"{o}dinger picture state.

\subsection{Measures of Entanglement}

When dealing with pure states, the Entropy of Entanglement (EOE) quantifies the amount of entanglement \cite{EntIntroPV}.  The EOE is defined to be the
von Neumann entropy of the reduced density matrix of the
system.\begin{equation}
  \mathcal{E}_{vN}(\hat{\chi})=S(Tr_{F}[\hat{\chi}])=S(\hat{\rho}_{A})=-\sum_{j}p_{j}\log_{2}(p_{j})\label{eq:EOE}\end{equation}
where the $p_{j}$ represent the eigenvalues of the reduced density matrix
$\hat{\rho}_{A}$. The EOE does not depend on which subsystem is traced
over. We will always trace over the field degrees of freedom.

When the total state of the system is mixed, the EOE can no longer be
used as an entanglement measure. While many measures of entanglement
have been defined, including the Entanglement of Formation, the
Distillable Entanglement, the Entanglement Cost, the I-Concurrence,
and others, most of these involve an infimum or supremum over a large
set (often the set of projector decompositions of $\hat{\chi}$) that is
in general difficult or impossible to evaluate analytically and quite
costly to compute numerically \cite{EntIntroPV}. We choose to focus on
an entanglement monotone known as the Logarithmic Negativity (LN)
\cite{NegVW} because we can
give explicit expressions for the value, and it can be easily
calculated numerically \protect \footnote{However, the LN only quantifies
  the degree to which the state in question violates the
  Peres-Horodecki positive partial transpose condition, and it is
  known that for bipartite quantum systems with a total Hilbert space
  dimension greater than six there exist entangled states, known as bound
  entangled states, that have positive partial transpose and,
  therefore, a LN of zero \cite{HHH}.%
}.

The LN is defined as \begin{equation}
  \mathcal{E_{N}}(\hat{\chi})\equiv\log_{2}\left\Vert
    \hat{\chi}^{T_{B}}\right\Vert _{1}\label{eq:LN}\end{equation}
where $\hat{\chi}^{T_{B}}$ is the partial transpose in the basis B of
the density matrix and \begin{equation} \left\Vert \hat{O}\right\Vert
  _{1}\equiv
  Tr\left(\sqrt{\hat{O}\hat{O}^{\dagger}}\right)\label{eq:1Norm}\end{equation}
is the trace norm of the operator $\hat{O}$. The partial transpose in
the basis $B=\{|j\rangle_{A}|k\rangle_{F}\}$ is defined such that
if\begin{equation}
  \hat{\chi}=\sum_{j,k,l,m}\chi_{j,k,l,m}|j\rangle\langle
  l|_{A}\otimes|k\rangle\langle
  m|_{F}\label{eq:chidecomp}\end{equation} then \begin{equation}
  \hat{\chi}^{T_{B}}=\sum_{j,k,l,m}\chi_{j,k,l,m}|j\rangle\langle
  l|_{A}\otimes|m\rangle\langle k|_{F}\label{eq:PT}\end{equation} so
that the second subsystem has been transposed. While the operator
produced by this procedure will depend on which subsystem is
transposed and in which basis, the trace norm depends only on the
eigenvalues, which will be independent of these choices \cite{EntIntroPV}.

In the case that the total state of the system is pure, the amount of
entanglement can be determined from the reduced density matrix for one
of the subsystems, as is clear from the definition of the EOE. In this
case one may show, using the Schmidt decomposition, that the LN
becomes\begin{equation}
  \mathcal{E_{N}}=\log_{2}\left(\left(\sum_{j}\sqrt{p_{j}}\right)^{2}\right)\label{eq:LNPure}\end{equation}
again with $p_{j}$ as the eigenvalues of the reduced density matrix.
When the total state of the system is mixed, however, knowing only the
reduced density matrix is not sufficient (since, for example, the
reduced density matrix of a maximally entangled state is the same as
when the total system is in a separable, completely-mixed state).
Thus, if we wish to track the evolution of entanglement for a total
mixed state, we must keep more information than just the reduced
density matrix.

\section{Entanglement Evolution}

\subsection{Methods of Solution\label{sec:SolMethods}}

For M=1 we have the Jaynes-Cummings model, and one can exactly
diagonalize $\hat{H}_{I}$ in the basis known as the dressed states.
The evolution of $\hat{\chi}$ in this basis is quite simple, and
knowing $\hat{\chi}(t)$ we can then calculate the entanglement using
the LN in general and the EOE if $\hat{\chi}(0)$ is a pure state.
For more than one EMF mode, one can still diagonalize the
Hamiltonian in the same way. Our initial state has all the
electromagnetic modes in their vacuum, so the energy arises from the
atomic state. This confines the initial state to a subspace spanned
by the zero- and one-excitation energy eigenstates of $\hat{H}_{0}$.
Because the total system evolves via a time-independent Hamiltonian
and $[\hat{H}_{0},\hat{H}_{I}]=0$, we know that the system will
remain in this subspace. We then only need to diagonalize
$\hat{H}_{I}$ in the subspace of degenerate one-excitation
eigenstates of $\hat{H}_{0}$. When there are $M$ field modes, this
subspace has dimension $M+1$. While this diagonalization is
straightforward numerically, doing this analytically requires the
solution to a polynomial of degree $M+1$, so for $M>3$ there is in
general no closed form solution.  In any case, having obtained the
eigenvalues the solution is otherwise exact.

For a pure initial state $\hat{\chi}(0)$, however, we only need the
reduced density matrix $\hat{\rho}_{A}$ to find the entanglement.
Anastopoulos and Hu \cite{AH} found the solution for the evolution of
$\hat{\rho}_{A}(t)$ in these types of systems with these initial
conditions. Given an initial atomic state $(1,\theta,\phi)$ at time
$t$ the reduced density matrix of the atom interacting with $M$ modes
is given by \begin{equation} \hat{\rho}_{A}(t)=\left(\begin{array}{cc}
      \frac{1}{2}u^{*}u(1+\cos(\theta)) &
      \frac{1}{2}ue^{i\phi}\sin(\theta)\\
      \frac{1}{2}u^{*}e^{-i\phi}\sin(\theta) &
      1-\frac{1}{2}u^{*}u(1+\cos(\theta))\end{array}\right)\label{eq:rhoA}\end{equation}
where \begin{equation}
  u(t)=\mathcal{L}^{-1}\left(\frac{1}{z+i\omega_{0}+\widetilde{\mu}(z)}\right)=\frac{1}{2\pi
    i}\int_{c-i\infty}^{c+i\infty}\frac{e^{zt}}{z+i\omega_{0}+\widetilde{\mu}(z)}dz,\label{eq:u}\end{equation}
$\mathcal{L}^{-1}$ represents the inverse Laplace transform,
and\begin{equation}
  \widetilde{\mu}(z)=\sum_{j=1}^{M}\frac{g_{j}^{2}}{z+i\omega_{j}}.\label{eq:mu}\end{equation}
The entanglement between the atom and the field modes depends only on
the eigenvalues of $\hat{\rho}_{A}(t)$,\begin{equation}
  p=\frac{1}{2}\left(1\pm\sqrt{1-4\cos^{4}\left(\frac{\theta}{4}\right)\left(\left|u\right|^{2}-\left|u\right|^{4}\right)}\right),\label{eq:eigvals}\end{equation}
so it depends only on the norm of $u(t)$. For $M$ field modes
\begin{equation}
  \left|u(t)\right|=\left|\sum_{j=0}^{M}\frac{\prod_{k=1}^{M}\left(z_{j}+i\delta_{k}\right)}{\prod_{l\ne
        j}^{M}\left(z_{j}-z_{l}\right)}e^{z_{j}t}\right|\label{eq:unorm}\end{equation}
with the $z_{j}$ values being all the solutions to the
equation\begin{equation}
  z\prod_{k=1}^{M}\left(z+i\delta_{k}\right)+\sum_{k=1}^{M}\left(g_{k}^{2}\prod_{l\ne
      k}^{M}\left(z+i\delta_{l}\right)\right)=0.\label{eq:pole}\end{equation}
Using Eq. (\ref{eq:eigvals}) we can compute the EOE or the LN. The
LN takes the particularly simple form\begin{equation}
  \mathcal{E_{N}}=\log_{2}\left(1+2\cos^{2}\left(\frac{\theta}{2}\right)\sqrt{|u|^{2}-|u|^{4}}\right)\label{eq:AHLN}\end{equation}

\subsection{Dependence on Initial Conditions}

In examining the evolution of entanglement in this system, one of the
primary questions is how it depends on initial conditions. Given the
Hamiltonian and the class of initial states we are considering, the
entanglement at any point in time is completely independent of the
parameter $\phi$ on the Bloch sphere for the initial state. While
this is clear for pure states from the form of Eq.
(\ref{eq:eigvals}), we show this is true for mixed initial states as
well. The unitary operator $\hat{U}_{\phi}(\alpha)\equiv
e^{-i\hat{H}_{0}\alpha/(\hbar\omega_{0})}$ acting on an initial state
of the form Eq. (\ref{eq:init}) will simply shift the angle $\phi$ of
the atomic state by an amount $\alpha$. Consider two initial states,
$\hat{\chi}(0)$ and
$\hat{\chi}'(0)\equiv\hat{U}_{\phi}(\alpha)\hat{\chi}(0)\hat{U}_{\phi}(\alpha)^{\dagger}$,
which differ only in their coordinate $\phi$ on the Bloch sphere.
Since $\hat{H}_{0}$ commutes with $\hat{H}_{I}$, we know
\begin{equation}
  \hat{\chi}'(t)=e^{-i\hat{H}_{I}t/\hbar}\hat{U}_{\phi}(\alpha)\hat{\chi}(0)\hat{U}_{\phi}(\alpha)^{\dagger}e^{i\hat{H}_{I}t/\hbar} 
  =\hat{U}_{\phi}(\alpha)e^{-i\hat{H}_{I}t/\hbar}\hat{\chi}(0)e^{i\hat{H}_{I}t/\hbar}\hat{U}_{\phi}(\alpha)^{\dagger}=\hat{U}_{\phi}(\alpha)\hat{\chi}(t)\hat{U}_{\phi}(\alpha)^{\dagger}.\label{eq:chiprime}
\end{equation}
Because $\hat{U}_{\phi}(\alpha)$ represents a local operation, we also know that
the degree of entanglement in the states $\hat{\chi}(t)$ and
$\hat{\chi}'(t)$ must be the same.

For the case where the initial state of the system is a pure state, we
can additionally conclude that, while the value of the entanglement at
a given time depends on the $\theta$ coordinate of the initial state,
qualitative features like the times at which local maxima and minima
of entanglement occur will be the same for all such initial pure
states. The easiest way to see this is by looking at Eq.
(\ref{eq:AHLN}), since the value of $\left|u(t)\right|$ at which the
LN reaches extrema does not depend on $\theta$.

\section{Specific Case Study}

\subsection{Single Mode --- The Jaynes-Cummings Model}

\begin{figure}
  \subfigure[$r=1$]{\includegraphics[bb=88bp 4bp 406bp 238bp,scale=0.5]{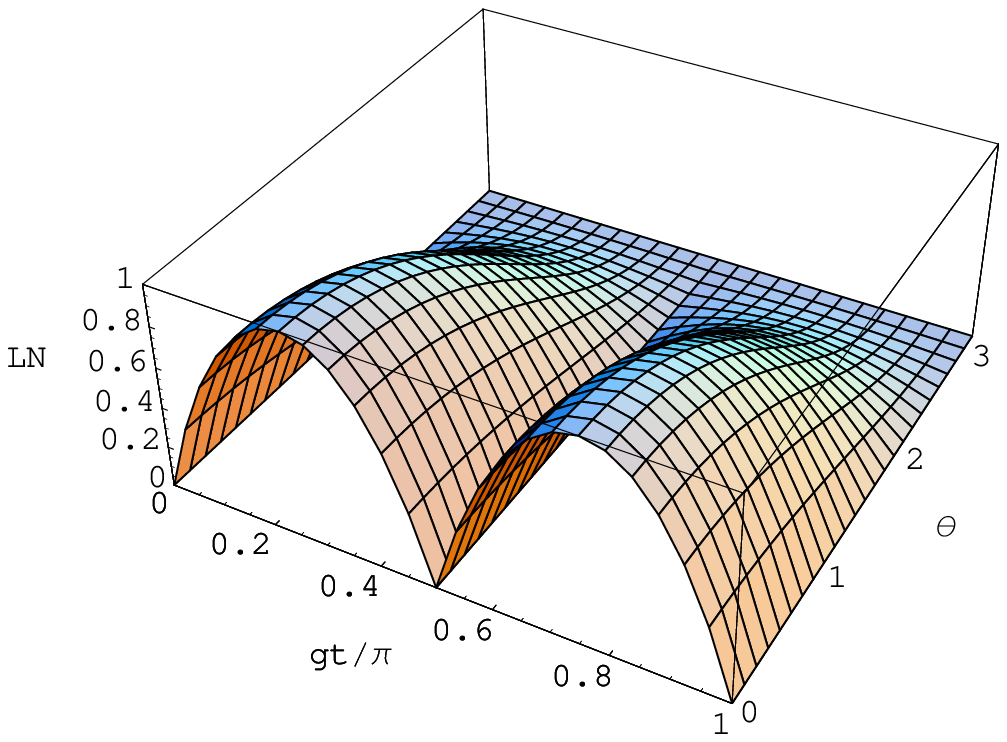}}\subfigure[$r=1/2$]{\includegraphics[bb=88bp 4bp 406bp 238bp,scale=0.5]{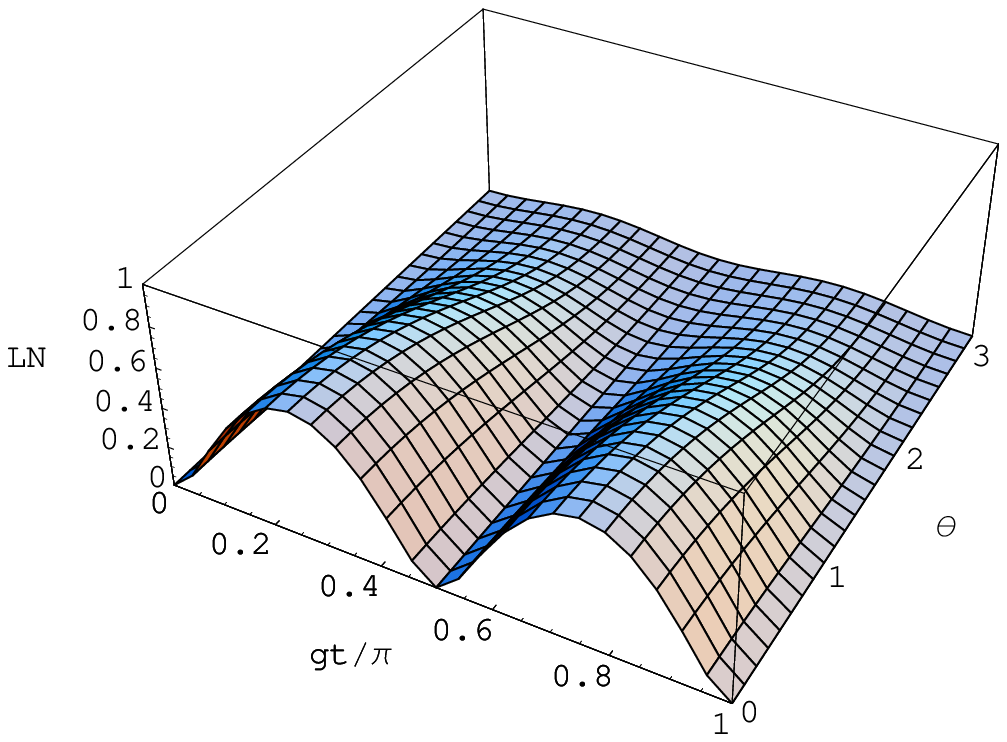}}\subfigure[$r=0$]{\includegraphics[bb=88bp 4bp 406bp 238bp,scale=0.5]{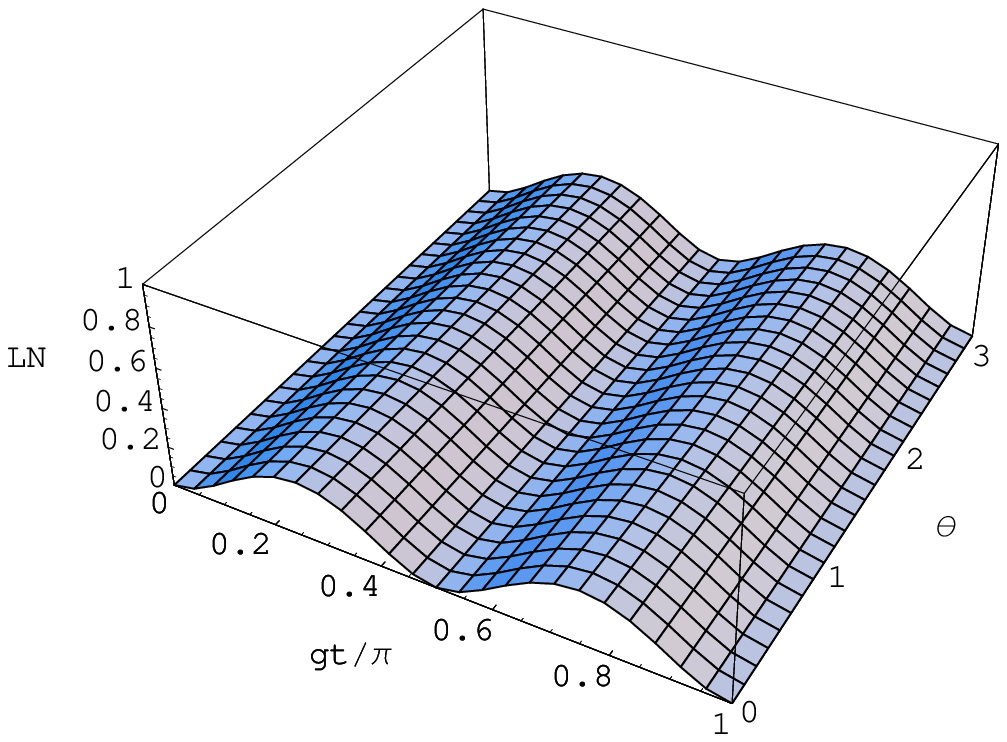}}

  \caption{\label{fig:initcond}LN as a function of time $t$ and the 
    initial state value of $\theta$ on the Bloch sphere for states with different values of $r$, where $g$ is the value of the atom-field coupling. In the last case, of course, all values of $\theta$ correspond to the same state.}
\end{figure}

We begin with the case of a single mode of the field ($M=1$)
interacting with the atom in the familiar Jaynes-Cummings model. The
Hamiltonian now depends only on the atom-field coupling $g$, the field mode frequency $\omega$, and the atomic transition frequency $\omega + \delta$. 
First, consider a system in which the atom is initially in a pure,
excited state and the field mode is resonant with the atomic
transition. In this case, the system undergoes Rabi oscillations
between the separable states $|e\rangle|0\rangle$ and
$|g\rangle|1\rangle$, passing through a maximally entangled state
each way. As a result, the entanglement has the relatively simple
oscillatory behavior with a time scale set by $g$.  For an initial pure state we have
\begin{equation}
u(t)=e^{-i(\omega + \frac{\delta}{2})t}
\left[ 
\left( \frac{\Delta\alpha + \delta}{2\Delta\alpha} \right)e^{-i\frac{\Delta\alpha}{2}t}
+ \left( \frac{\Delta\alpha - \delta}{2\Delta\alpha} \right)e^{i\frac{\Delta\alpha}{2}t}
\right]
\end{equation}
where $\Delta\alpha \equiv \sqrt{\delta^{2}+4g^{2}}$.  For initially mixed atomic states, the fully density matrix for the bipartite system may be computed using the dressed states, as discussed in Sect. \ref{sec:SolMethods}. The variation in the behavior of entanglement in the $\delta=0$ case
for different initial conditions is shown in Fig. \ref{fig:initcond}.%
As stated above it does not depend on $\phi$, and many of the
qualitative features such as the times at which maximal and minimal
entanglement occur do not depend on $\theta$.%

\begin{figure}
  \includegraphics[scale=0.7]{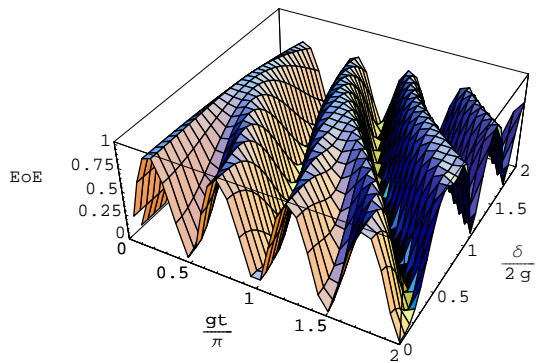}

  \caption{\label{fig:detuningeffect}The dependence of entanglement on
    the parameters of the Hamiltonian, the detuning $\delta$ and atom-field coupling $g$. 
    The odd numbered minima in the
    entanglement (as a function of time $t$) increase and eventually
    disappear as the detuning increases.}
\end{figure}

The other question is how the entanglement generated by the dynamics
of the system depends on the parameters of the Hamiltonian. As is well
known for the Jaynes-Cummings model, the time scale of the evolution
is the vacuum effective Rabi frequency
$\frac{1}{2}\sqrt{\delta^{2}+4g^{2}}$ . What's more interesting is the
dependence on the detuning $\delta$ shown in Fig. \ref{fig:detuningeffect}.%

\begin{figure}
  \includegraphics[scale=0.4]{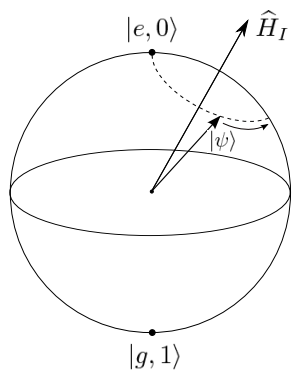}

\caption{\label{fig:blochsphere}The Bloch sphere for the one-excitation manifold $\hat{H_{0}}$. 
The vector representing the quantum state of the system $|\psi\rangle$
precesses in a circle about the vector for the Hamiltonian $\hat{H}_{I}$ as time
advances. When the state vector encounters either pole of the
sphere, the state becomes separable. If the state vector crosses the
equator of the sphere, the state is maximally entangled at that
instant. The ratio of the detuning $\delta$ to the atom-field coupling $g$ determines the angle the
Hamiltonian vector makes with the vertical axis. If $\delta=0$ the
Hamiltonian vector is horizontal, and as $\delta/g$ becomes large
the Hamiltonian vector approaches the vertical. The vector for the
state starts on the pole of the sphere (due to the initial
condition) and travels around a circle centered on the Hamiltonian vector; thus, the circle of precession is a great circle when
$\delta=0$, and the circle becomes very small when $\delta/g$ is
large.}
\end{figure}
For an initial pure state, we can understand the time evolution of
entanglement by focusing on the two level system composed of the
one-excitation states of $\hat{H}_{0}$. This two level system can
again be mapped to the Bloch sphere. As is well known, any
time-independent Hamiltonian acting on this system can be
represented by a fixed vector, and the state's evolution under the
Hamiltonian can simply be described by a precession of the Bloch
vector for the state about the Hamiltonian vector. If we choose the
basis to be $B=\{|e\rangle|0\rangle,|g\rangle|1\rangle\}$ where the
first state lies at the top of the Bloch sphere and the second state
lies at the bottom, then we have\[
\hat{H}_{I}=\hbar\left(\begin{array}{cc}
-\delta/2 & g\\
g &
\delta/2\end{array}\right)=-\frac{\delta}{2}\hat{\sigma}_{z}+g\hat{\sigma}_{x}=(g,0,-\frac{\delta}{2})\cdot\vec{\sigma}\]
where $\vec{\sigma}$ is the (pseudo) vector associated with the
Pauli matrices, and $(g,0,-\frac{\delta}{2})$ becomes the vector
representing the Hamiltonian. Notice that all points along the
"equator" of this Bloch sphere are maximally entangled, because
they are all equivalent to
$\frac{1}{2}(|e\rangle|0\rangle+|g\rangle|1\rangle)$ by a local
unitary operation. When the detuning is small, the vector for
$\hat{H}_{I}$ points almost along the x-axis, and the state rotates
from $|e\rangle|0\rangle$ to $|g\rangle|1\rangle$ as it evolves in
time, crossing the equator twice, so it passes through two maximally
entangled states and two separable states during each rotation. Fig.
\ref{fig:blochsphere} illustrates this Bloch sphere picture of the
entanglement evolution, which accounts for all the qualitative
features that appear in Fig. \ref{fig:detuningeffect}.

\subsection{Multiple Modes}

\begin{figure}
\includegraphics[scale=0.7]{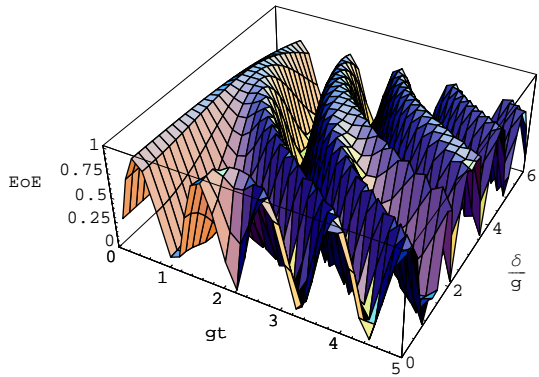}

\caption{\label{fig:twomodesymparamdep}The dependence of the EOE on both time t and the detuning
$\delta$ from the atomic resonance frequency for the case where there are two
EMF modes.}
\end{figure}

With additional modes of the electromagnetic field, the behavior of
the system quickly becomes more complex. Let us begin with the
bipartite entanglement between the atom and the EMF when the
field has only two modes. First consider the {}``symmetric'' case,
where $g_{1}=g_{2}\equiv g$ and $\delta_{1}=-\delta_{2}\equiv\delta$.
If the initial state is pure, then we have\[
u(t)=\frac{\delta^{2}+2g^{2}\cos\left(t\sqrt{2g^{2}+\delta^{2}}\right)}{\delta^{2}+2g^{2}}\]
 and we can easily compute the entanglement as a function of time.
If the initial atomic state is mixed, we can still compute the LN by the technique mentioned in \ref{sec:SolMethods}. 
The dependence on initial conditions in Fig. \ref{fig:twomodesymparamdep} appears quite similar
to the single mode case, while the dependence on $\delta$ shown in
Fig. \ref{fig:twomodesymparamdep} is somewhat more complex. We are more interested, however, in cases where one of the field modes is very close to resonance, so a more practical case to investigate is that of three field modes.%

In considering a finite number of field modes, we are primarily thinking
of modeling the intracavity field in cavity QED, particularly in the
optical regime. We will adopt the simplest possible model for such
a cavity, two perfectly conducting, infinite, parallel plates(the cavity 
model used in \cite{AH}). There are two
classes of modes between the plates: the discrete set of modes with
a purely longitudinal wave vector, which we will call the "cavity
modes", and the continuum of modes with a transverse component,
which propagate outward along the plates. If we ignore the coupling
between the atom and these transverse traveling modes but rely on
the fact that the effect of a single mode diminishes as it becomes
detuned far from resonance, then we can try to approximate the cavity
by only the finite number of modes $M$. This should reflect an actual
cavity in the strong-coupling limit, where $g$ is much larger than
the spontaneous emission rate or the cavity decay rate.

Considering the frequencies of these cavity modes, if the mode
closest to resonance is detuned by an amount $\delta$ then all other
modes will be detuned by an amount $\delta+k\Delta$ where $\Delta$ is
the free spectral range of the cavity in angular frequency, which is 
$\Delta=\frac{\pi c}{L}$ for our simple model.  As noted earlier, all
the $g_{j}$ values are proportional to each other, so if we call the
value for the mode closest to resonance $g$, this sets the value for
all the other coupling constants. Thus, we are left with only four
constants whose values need be specified.

We will consider the regime where $\delta\ll\Delta$ (as is often
the case in cavity QED experiments); thus, one mode is almost in resonance,
and the next nearest two modes have almost the same detuning. As a
result, it will make sense to consider only odd total numbers of modes
$M$ in trying to understand the cavity. With this further specificity,
we may rewrite the formula for $\left|u(t)\right|$. Let $M=2Q+1$,
then \begin{equation}
\left|u(t)\right|=\left|\sum_{j=0}^{M}\frac{\prod_{n=1}^{Q}\left(w_{j}^{2}+n^{2}\Delta^{2}\right)}{\prod_{l\ne j}^{M}\left(w_{j}-w_{l}\right)}e^{w_{j}t}\right|\label{eq:ushifted}\end{equation}
 where $w_{j}$ are solutions to the equation 
\begin{equation}
\left(w^{2}-i\delta w+g^{2}\right)\prod_{n=1}^{Q}\left(w^{2}+n^{2}\Delta^{2}\right)+ \\
2wg^{2}\sum_{n=1}^{Q}\left(1-\frac{n^{2}\Delta^{2}}{\left(\omega_{0}+\delta\right)^{2}}\right)^{-1} \left(w+\frac{in^{2}\Delta^{2}}{\left(\omega_{0}+\delta\right)}\right)\prod_{j\ne n}^{Q}\left(w^{2}+j^{2}\Delta^{2}\right)=0.\label{eq:shiftedpole}
\end{equation}

\begin{figure}
\subfigure[]{\includegraphics[scale=0.75]{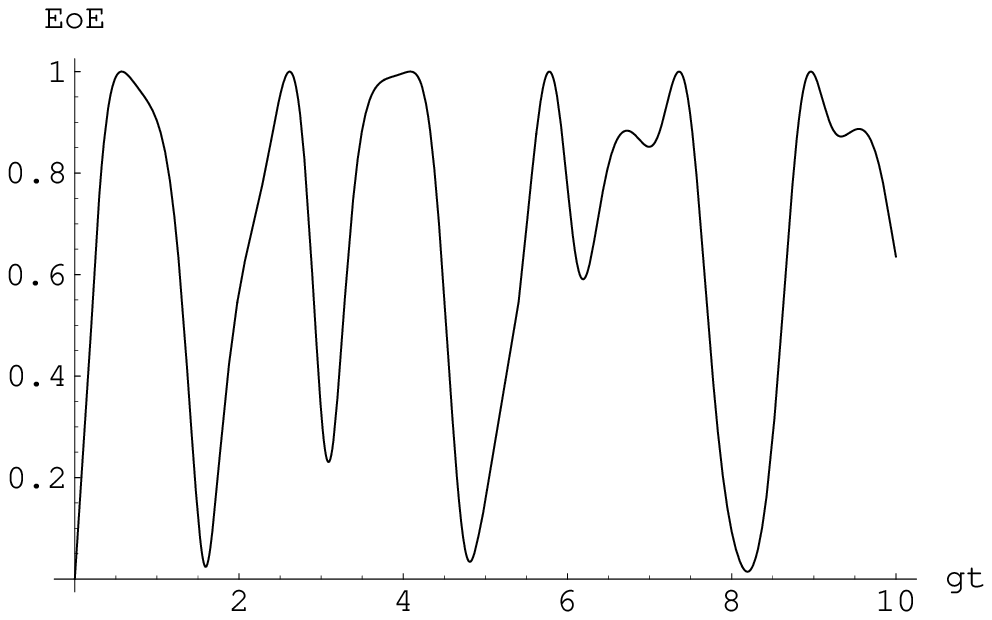}}
\subfigure[]{\includegraphics[scale=0.75]{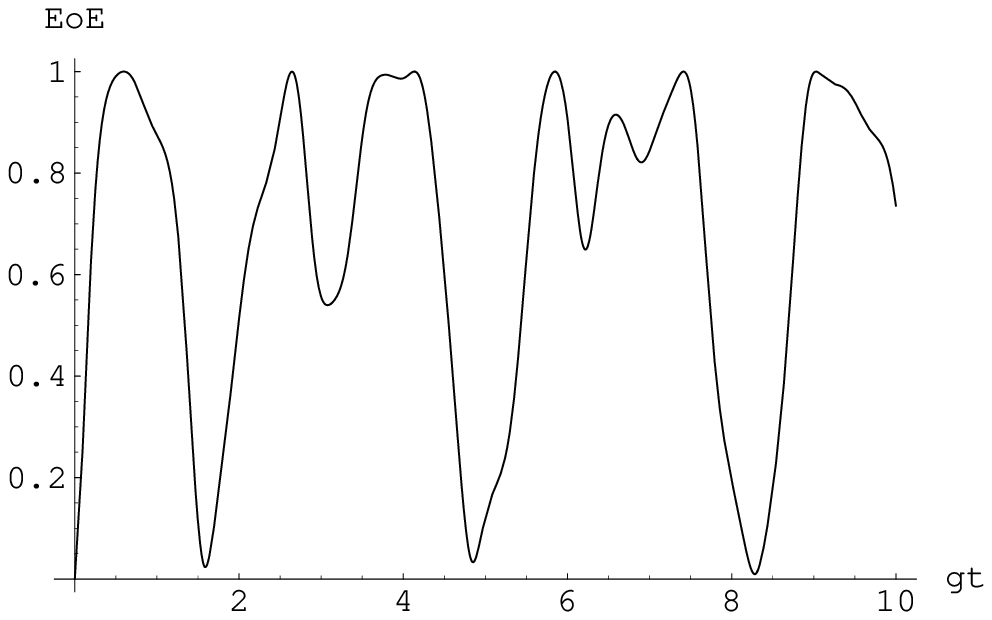}}

\caption{\label{fig:Mmodeevolunphys}The evolution with time t of the EOE
between an atom in an initially pure, excited state and (a) three
modes or (b) five modes of the EMF. To highlight the effect of
additional modes, we choose the values for the detuning $\delta$, atom-field coupling $g$, and free spectral range $\Delta$ so that  $\Delta/g=5$, $g/\delta=10$,
and $\omega_{0}/g=10^{7}$, though these are unlikely to be
characteristic of experimental values.}
\end{figure}

We consider first the $M=3$ case. When $\Delta$ is small, the time
evolution behavior, shown in Fig. \ref{fig:Mmodeevolunphys}(a), becomes
considerably more complex with the contributions of more frequency
components to $u(t)$. The effect of the addition of two more modes
in the $M=5$ case is shown in Fig. \ref{fig:Mmodeevolunphys}(b).
The clear intuitive interpretation that was present in the single
mode case is not obvious in these cases. %

In optical frequency cavity QED experiments, one is generally working
in the regime where $\omega_{0}\gg\Delta\gg g$ and $\delta$ is smaller
than or on the same order as $g$. In this case, other cavity modes
are far enough away in frequency to have little effect on the atom,
and the evolution for the multi-mode models becomes nearly identical
to that of the single mode case. We can calculate the poles in this
regime by writing Eq. (\ref{eq:shiftedpole}) in terms of the dimensionless
parameters $g/\Delta$, $\Delta/\omega_{0}$, and $\delta/\omega_{0}$
and solving it perturbatively in these small parameters. In experiments,
these parameters might typically have values on the order of $10^{-4}$,
$10^{-3}$, and $10^{-7}$ respectively. This perturbative analysis
leads to the poles\begin{align}
w &=-\frac{i}{2}\left(\delta\pm\sqrt{\delta^{2}+4g^{2}}\right) \label{eq:singularpole}\\
w &=\pm i\left(k\Delta+\frac{g^{2}}{k\Delta}\right)\qquad k\in\left\{ 1,2,...,\frac{M-1}{2}\right\} \label{eq:otherpoles}
\end{align}
keeping up to terms linear in $\delta/\omega_{0}$ and terms quadratic
in $g/\Delta$ and $\Delta/\omega_{0}$ (including terms of order
$g/\omega_{0}$). This is the lowest order at which the poles in Eq.
(\ref{eq:otherpoles}) yield non-vanishing contributions to $|u(t)|$ %
\protect \footnote{If one keeps only to zeroth order in $\delta/\omega_{0}$ and linear
order in $g/\Delta$ and $\Delta/\omega_{0}$, then the resulting
expression for $|u(t)|$ is identical to the single-mode case, as
one may expect when all other modes are very far off resonance.%
}. In the sum from Eq. (\ref{eq:ushifted}), the poles of the form shown
in Eq. (\ref{eq:singularpole}) give terms of order unity, while poles
of the form shown in Eq. (\ref{eq:otherpoles}) yield terms of order
$g^{2}/\Delta^{2}$, so in this regime the effects of additional
modes are small, but we have found an analytic expression for the
behavior of the system for an arbitrary, finite number of modes.

\subsection{Full Intracavity Field}

Another approach toward understanding the effect of the cavity is to
look at the long-time behavior of entanglement in the presence of the
infinite field modes of the cavity. Again, we will use the simple
cavity model mentioned earlier consisting of two infinitely large,
perfectly conducting parallel plates. We can begin with a box having
a square transverse cross-section of area $A$ and a longitudinal
length $L$ with boundary conditions such that the field vanishes on
the plates at the longitudinal boundaries and is periodic on the
transverse boundaries. In this case we can write
$g_{\vec{k}}=\frac{\lambda}{\sqrt{\omega_{\vec{k}}LA}}$, where the
dimensionless quantity $\lambda$ is the strength of the coupling to
the overall field. As found in \cite{AH},
\[
\mu(s)=\sum_{\vec{k}}g_{\vec{k}}^{2}e^{-i\omega_{\vec{k}}s}\rightarrow\frac{\lambda^{2}}{2\pi^{2}L}\sum_{n=-\infty}^{\infty}\int_{\left|\frac{\pi n}{L}\right|}^{\infty}e^{-ik^{\prime}s}dk^{\prime}
\]
in the continuum limit where $A\rightarrow\infty$.  Since this
integral clearly diverges,  we add an exponential cutoff by taking
$s\rightarrow s-i\epsilon$ to regularize it. With this cutoff
\begin{equation}
\mu_{\epsilon}(s)=\frac{\lambda^{2}}{2\pi^{2}L}\sum_{n=-\infty}^{\infty}\int_{\left|\frac{\pi
n}{L}\right|}^{\infty}e^{-ik^{\prime}s-k^{\prime}\epsilon}dk^{\prime}=\frac{\lambda^{2}}{\pi
L}\frac{1}{\epsilon+is}\frac{1+e^{-i\pi(s-i\epsilon)/L}}{1-e^{-i\pi(s-i\epsilon)/L}}.\label{eq:cavitymu}\end{equation}
Taking the Laplace transform yields approximately \cite{AH}
\begin{equation}
\widetilde{\mu_{\epsilon}}(z)=\frac{-i\lambda^{2}}{\pi^{2}\epsilon}+\frac{\lambda^{2}}{\pi^{2}}z \ln\left(ie^{\gamma_{e}}\epsilon z\right) - \frac{i\lambda^{2}}{\pi L}\left[\ln\left(\Gamma\left(\frac{Lz}{i\pi}\right)\right)-\frac{Lz}{i\pi}\ln\left(\frac{Lz}{i\pi}\right)+\frac{Lz}{i\pi}+\frac{1}{2}\ln\left(\frac{Lz}{2i\pi^{2}}\right)\right]+\mathcal{O}(\epsilon)\label{eq:cavitymutilde}
\end{equation}
where $\ln\left(\Gamma(z)\right)$ is defined such that it has a branch
cut on each ray $\{-n+iy\,|\,n\in\mathbb{N}\,\&\,y\in\mathbb{R}\geq 0
\}$, and $\gamma_{e}$ is the Euler-Mascheroni constant. Clearly this
expression depends on the cutoff $\epsilon$, which we assume to be
small. One cannot in the end take the limit of $\epsilon\rightarrow
0$ for this model, even with renormalization of the model parameters.
$\epsilon$ should be regarded as a phenomenological parameter that
reflects the fact that at high frequencies the approximations that
underlie our model (the rotating wave approximation, the two level
approximation, etc.) must give way to other physics.

Inserting this result for $\widetilde{\mu}(z)$ into Eq. (\ref{eq:u}), $u(t)$ may be obtained by
finding the residues arising from the poles of the integrand. As found in \cite{AH}, all poles have a negative real part. From Eq. (\ref{eq:unorm}) one can see that the negative real part of each
pole will cause its contribution to $u(t)$ to die off exponentially
with time, so the pole with the greatest real part will dominate at long
times.

If $\lambda$ is sufficiently small, this pole can be found
by doing a perturbative expansion in $\lambda$ that yields
\begin{equation}
z_{p}=-i\widetilde{\omega}_{0}-\widetilde{\mu_{\epsilon}}(-i\widetilde{\omega}_{0})+\mathcal{O}\left(\lambda^{4}\right)=-i\Omega-\gamma\label{eq:PerturbPole}
\end{equation}
where $\widetilde{\omega}_{0}\equiv\omega_{0}-\frac{\lambda^{2}}{\pi^{2}\epsilon}$
for notational convenience. Let us further define \begin{equation}
\Omega_{\infty}\equiv\widetilde{\omega}_{0}-\frac{\lambda^{2}\widetilde{\omega}_{0}}{\pi^{2}}\ln\left(e^{\gamma_{e}}\epsilon\widetilde{\omega}_{0}\right).\label{eq:OmegaInf}\end{equation}
$\Omega_{\infty}$ would be the value of $\Omega$ in the case of the free field and, therefore, the physically observable dressed value of the two-level transition frequency.  If we rewrite our expression for $z_{p}$ in terms of this dressed atomic frequency, we may write \begin{equation}
\widetilde{\omega}_{0}=\Omega_{\infty}+\frac{\lambda^{2}\Omega_{\infty}}{\pi^{2}}\ln\left(e^{\gamma_{e}}\epsilon\Omega_{\infty}\right)+\mathcal{O}\left(\lambda^{4}\right),\label{eq:omega Expansion}
\end{equation}
so then the pole from Eq. (\ref{eq:PerturbPole}) becomes
\begin{equation}
z_{p}=-i\Omega_{\infty}-\frac{i\lambda^{2}}{\pi L}\left\{ \ln\left[\Gamma\left(-\frac{L\Omega_{\infty}}{\pi}\right)\right]+\frac{L\Omega_{\infty}}{\pi}\ln\left(-\frac{L\Omega_{\infty}}{\pi}\right)-\frac{L\Omega_{\infty}}{\pi}+\frac{1}{2}\ln\left(-\frac{L\Omega_{\infty}}{2\pi^{2}}\right)\right\}
 +\mathcal{O}\left(\epsilon\right)+\mathcal{O}\left(\lambda^{4}\right).\label{eq:ApproxCavityPole}
\end{equation}
This perturbative solution will be valid as long as $\lambda$ is sufficiently small that the higher order terms can be ignored and no branch cut lies between $-i\Omega_{\infty}$ and the pole calculated in Eq. (\ref{eq:ApproxCavityPole}).  Clearly, the expression will not be valid when $L\Omega_{\infty}/\pi$ is a non-negative integer, which is the condition for resonance. When $\Omega_{\infty}$ is sufficiently close to resonance, we can verify numerically that there will be two closely spaced poles, as one would expect from a simple Jaynes-Cummings model treatment.

In the case that the perturbative solution is valid and gives the only pole important at long times, then in that limit $|u(t)|\propto e^{-\gamma t}$ from Eq. (\ref{eq:PerturbPole}). The value of $\gamma$ depends on the value of the quantities $\Omega_{\infty}$, $\lambda$, $\epsilon$, and $L$.  When $\lambda$ is sufficiently small, $\gamma$ may be obtained by taking the real part of Eq. (\ref{eq:ApproxCavityPole}).  In other cases it may be found numerically.  Having determined the poles, if they are first order then the residue for each will simply be $\left( 1 + \frac{d\widetilde{\mu}(z)}{dz}\bigr\rvert_{z_{p}} \right)^{-1}$. Combining the expression for $|u(t)|$ at long times with our expressions for entanglement of a pure state yields a relatively simple behavior for the entanglement.  The exponential decay of atomic coherence noted by \cite{AH} leads to entanglement that falls off toward zero at long times.%

\section{Summary and Discussion}

In this work we have considered the interaction between a two-level
atom and different numbers of electromagnetic field modes with the
aim of gaining a detailed description and deeper understanding of the
nature and dynamics of quantum entanglement between the atom and the
field. We have produced exact results (given the usual atom-field interaction Hamiltonian 
derived from making the rotating wave, dipole, and
two-level approximations) that go beyond the usual single-mode
Jaynes-Cummings model and give the complete development of
quantum entanglement in time for a system where the field is
initially in the vacuum state.

On the effect of initial conditions of the atom in the general case,
we find that quantum entanglement is not affected by the $\phi$ angle
of the initial atomic state on the Bloch sphere.  For initially pure
atomic states, we also find that qualitative features of the time
evolution of entanglement remain the same for different values of
$\theta$.

For the case where the cavity has only one dominant field mode, our
calculations reproduce the familiar results for the Jaynes-Cummings
model obtained before \cite{Gea90,Gea93,PhoKni}. Our result shows the
expected oscillations of entanglement between the atom and the field,
including periodic complete disentanglement in the resonant case.
While many prior treatments assume a pure initial atom-field state
or that the field mode is resonant with the 
2LA, we have calculated the entanglement when the initial state (and,
thus, the state at all times) of the atom-field system is mixed. We
have also shown the effects of detuning from resonance on the
dynamics, giving a simple conceptual picture that accounts for all
qualitative features.

For the situation when the atom is coupled to a finite number of
electromagnetic field modes, since our calculation made no weak-coupling
approximation, we are able to give a detailed account of the behavior
of entanglement in the strong-coupling regime. Indeed, if one wishes
to relate this to an actual cavity QED system where losses due to
spontaneous emission and cavity damping are present, our closed-system
treatment will be useful exclusively in the case of strong coupling
$g\gg\kappa,\gamma$ at sufficiently early times where $\kappa t\ll1$
and $\gamma t\ll1$, wherein dissipation is insignificant and can
be ignored.

We have discussed how the effects of these additional modes far from
resonance are small in the normal parameter range of current cavity
QED experiments. On the other hand, if the system were placed in the
super-strong-coupling regime, where $g$ is comparable to the free
spectral range of the cavity $\Delta$, then we can see from Fig.
\ref{fig:Mmodeevolunphys} that the dynamics are significantly altered
from those of the single-mode case. Our calculations add new, significant
features to the behavior of the entanglement.

It has been suggested \cite{MeiMey} that one may be able to reach the super-strong-coupling
regime experimentally in a system with many ultracold atoms trapped
in an optical lattice inside the cavity. Under the right circumstances,
our calculations might apply to such a system. If the experimental realization
were such that the intracavity field coupled almost identically to
the internal states of all $N$ atoms so that the system behaved like
the Tavis-Cummings model \cite{TavisCummings}, then our model could
still be applied if the atoms were initially prepared in the Dicke
state $\left|N/2,-(N-1)/2\right\rangle $ \cite{Dicke} (i.e., a collective
state with one excitation symmetrized over all atoms). In this case,
the collective dipole of the atomic sample behaves like a single,
two-level atom with enhanced coupling constant $g\rightarrow g\sqrt{N}$.
Of course, in general the situation may be considerably more complex
than a Tavis-Cummings type model, but at the least our results suggest
interesting new dynamics for quantum entanglement in this super-strong-coupling regime.

Finally, we have treated the case of a 2LA interacting with the
infinite number of modes present in an intracavity field and given an account
of the long-time behavior of their entanglement. In this model we have assumed that
the mirrors are perfectly reflecting. There are two possible approaches to applying these results
to a physical cavity with damping. If the damping time scale is sufficiently longer than the other important dynamical time scales then the damping can be ignored for times $\kappa t\ll1$, and one can apply our results directly.  Alternately, one could
modify the treatment we have given by using cavity field modes that extend through the cavity mirrors to the exterior, as would exist
with partially transmitting mirrors, though the entanglement calculated would still be with the entire field, not merely the intracavity portion.

Though our solution for the cavity case could be evaluated numerically
at any time, we pay special attention to the long-time behavior where an analytic
result can be found. In order for this regime to be meaningfully applied
to a physical system with $\kappa\ne0$, one requires that such long
times are still significantly shorter than $1/\kappa$. The long-time
solution we have presented requires all other contributions to $u(t)$
to be negligibly small. These correspond to poles in the complex plane
with negative real parts that are larger in absolute value. If the
next most significant term has a real part $-\gamma^{\prime}$, then
the time scale that defines long times will be $\left(\gamma^{\prime}-\gamma\right)^{-1}$.
Given that we expect any other poles to move off to infinity in the
complex plane as $\lambda\rightarrow0$, if $\lambda$ is small enough
the time scale that defines long times should become comparatively short, and there should
be a window of validity where it can be considered long times and
yet $\kappa$ may be ignored.

In obtaining these exact solutions, we are able to see several
features of the dynamics.  With the infinite collection of modes, we
find that the terms from Eq. (\ref{eq:u}) are exponentially decaying,
leading to a behavior in which the entanglement dissipates at long
times.  By contrast, when only a few field modes are included each
term is purely oscillatory and no such decay is present.   In each of
the cases with an infinite collection of field modes, our exact
calculation leads to the expression for $\widetilde{\mu}(z)$ that can
be used to evaluate $u(t)$ numerically in order to obtain the
behavior of the system. To give more concrete results, we have
examined the long-time behavior working out the result perturbatively
in the coupling $\lambda$. While this long-time behavior results in
an exponential decay of coherence in the atom \cite{AH}
characteristic of Markovian behavior of the atom (now considered as
an open system with the field modes integrated away), our exact
result for the total system can provide a check on how good the
various approximations introduced for any description of the open
system could be.  These features could be explored by solving the
equations we derived here numerically in particular limits of
interest.

We addressed the strong-coupling regime above. Another important
feature which our study of the full dynamics of the total system can
provide some comparison with is the non-Markovian versus Markovian
dynamics of a related open system.  As is explained in the
Introduction a la Zwanzig's projection operator paradigm, the exact
and complete description of the total system (via an ordinary
differential equation) can be transcribed to that of a distinguished
subsystem (where one desires a detailed description) with the
dynamics of all the other subsystems it interacts with subsumed in
the non local kernels of integro-differential equation it obeys. This
is the origin of non-Markovian dynamics with memory in a closed
system. The open-system dynamics is only one step away from this,
when one decides to coarse-grain away to varying extents the degrees
of freedom of the other subsystems the distinguished system
interacts with. That is why the dynamics of such open systems are
generically non-Markovian dynamics. The unitary dynamics of a closed
combined system thus offers the best way to compare the relation of
the non-Markovian dynamics of the open system with the Markovian
dynamics of the same open system and assess the accuracy of various approximations
 introduced for its description. In the cases that
include a finite number of field modes, one can obtain such
subsystem dynamics quite directly from our solution for the total
system dynamics. For a long time quantum atomic-optical experiments
 have operated in parameter ranges where the dynamics give relatively
simple yet highly accurate results, so these issues have been largely
ignored. But with stronger couplings, shorter times, and many
correlated atoms present (the case of cold atoms in optical lattices
is a familiar one) this luxury may not be available for too long.

The ever increasing capability for exquisite control of atom-field
interactions in the laboratory and the increasing demands of quantum
information applications require ever increasing accuracy in our
theoretical modeling of these quantum systems.  In addition, optical
cavity QED is now truly approaching the regime of strong coupling. By
working out the dynamics of quantum entanglement in this simple
system exactly, in the regime of strong coupling and without making
any compromising approximations (like the popular Born-Markov
approximation), we hope to begin to uncover some new features of this
realm of physics. Ours is really just a small step in this direction.
There remains much work to be done, both to understand the generic
behaviors of these simple systems by model studies and to add more
features to the theoretical models that can provide a closer
depiction of reality, captured in the near future by higher precision
experiments.

\begin{acknowledgments}
 We are indebted to Charis Anastopoulos for very detailed and
 helpful discussions. NC thanks Luis Orozco and Perry Rice for useful
 comments. This work is supported in part by grants from
 NSA-LPS, NIST and NSF-ITR program (PHY-0426696).
\end{acknowledgments} 

\bibliographystyle{h-physrev}
\bibliography{pybdb}

\end{document}